\documentclass[AMA,STIX1COL]{WileyNJD-v2}
\articletype{}%

\received{XX.XX.XXXX}
\revised{YY.YY.YYYY}
\accepted{ZZ.ZZ.ZZZZ}

\usepackage{graphicx}  
\usepackage{booktabs}  

\providecommand{\expect}{\mathrm{E}}
\providecommand{\var}{\mathrm{Var}}
\providecommand{\normaldistn}{\mathrm{Normal}}
\providecommand{\unifdistn}{\mathrm{Uniform}}
\providecommand{\halfnormaldistn}{\mathrm{half\mbox{-}Normal}}
\providecommand{\mean}{\mu}
\providecommand{\vari}{\sigma^2}
\providecommand{\cv}{c_\mathrm{v}}

\begin{document}
  \title{Summarizing empirical information on between-study heterogeneity for Bayesian random-effects meta-analysis}
  \author[1]{Christian R\"{o}ver}
  \author[2]{Sibylle Sturtz}
  \author[2]{Jona Lilienthal}
  \author[2]{Ralf Bender}
  \author[1]{Tim Friede}
  \authormark{C.~R\"{O}VER, S. STURTZ, J. LILIENTHAL, R. BENDER, T.~FRIEDE}

  \address[1]{\orgdiv{Department of Medical Statistics}, \orgname{University Medical Center G\"{o}ttingen}, \orgaddress{\state{G\"{o}ttingen}, \country{Germany}}}
  \address[2]{\orgdiv{Department of Medical Biometry}, \orgname{Institute for Quality and Efficiency in Health Care (IQWiG)}, \orgaddress{\state{K\"{o}ln}, \country{Germany}}}
  \corres{*Christian R\"{o}ver, \email{christian.roever@med.uni-goettingen.de}}

\abstract[Summary]{
  In Bayesian meta-analysis, the specification of prior probabilities for the between-study heterogeneity is commonly required, and is of particular benefit in situations where only few studies are included. Among the considerations in the set-up of such prior distributions, the consultation of available \emph{empirical data} on a set of relevant past analyses sometimes plays a role. How exactly to summarize historical data sensibly is not immediately obvious; in particular, the investigation of an empirical collection of heterogeneity \emph{estimates} will not target the actual problem and will usually only be of limited use. The commonly used normal-normal hierarchical model for random-effects meta-analysis is extended to infer a heterogeneity prior. Using an example data set, we demonstrate how to fit a distribution to empirically observed heterogeneity data from a set of meta-analyses. Considerations also include the choice of a parametric distribution family. Here,  we focus on simple and readily applicable approaches to then translate these into (prior) probability distributions.}

\jnlcitation{\cname{\author{C. R\"{o}ver}, 
                    \author{S. Sturtz}, 
                    \author{J. Lilienthal}, 
                    \author{R. Bender}, 
                    \author{T. Friede}} 
             (\cyear{2022}), 
             \ctitle{Summarizing empirical information on between-study heterogeneity for Bayesian random-effects meta-analysis}, 
             \cjournal{(submitted for publication)}, \cvol{2022}.}
\maketitle

\section{Introduction}
  A range of statistical modelling approaches is available for random-effects meta-analyses; here we consider meta-analysis within the common and general framework of the \emph{normal-normal hierarchical model (NNHM)} where measurement uncertainty as well as be\-tween-study heterogeneity are modelled based on normal distributions.\citep{HedgesOlkin,Roever2020} 
  The NNHM provides a versatile framework for meta-analysis that is applicable in a wide range of cases where the data may be summarized by a point estimate along with a measure of uncertainty (a standard error or confidence interval). Examples include many types of endpoints, e.g., (standardized) mean differences (logarithmic) odds ratios, relative risks or hazard ratios, prevalences, correlation coefficients, and many more.
  Analysis based on the NNHM and performed within a Bayesian framework requires the assignment of prior distributions for the overall mean effect~($\mu$) and the heterogeneity standard deviation~($\tau$); while the former is commonly uncontroversial and a uniform (uninformative) specification if often appropriate, the latter may require more care, in particular in cases where only few studies are included in the meta-analysis.\citep{FriedeRoeverWandelNeuenschwander2017a,RoeverEtAl2021,BenderEtAl2018}

  Generally speaking, the construction of prior distributions may be approached from a range of different angles,\citep{Jaynes1968,KassWasserman1996,ZondervanZwijnenburgEtAl2017,BestDallowMontague2020} some of which include the consideration of empirical data. Besides the empirical evidence, a number of further aspects may determine details of the prior distribution's specification, such as tail behaviour, robustness or conservatism.\citep{Roever2020,RoeverEtAl2021} 
  Practical application eventually also requires priors to be reasonably simple, 
  and easily motivated, communicated and implemented.

  Several authors have investigated empirical data on heterogeneity in meta-analysis applications previously.\citep{HigginsWhitehead1996,Pullenayegum2011,TurnerEtAl2012,SteelEtAl2015,LanganHigginsSimmonds2015,vanErpEtAl2017,SeideEtAl2018b,SeideEtAl2019,GunhanRoeverFriede2020}
  In some instances, this resulted in a sample of heterogeneity \emph{estimates}, which may provide a rough idea of plausible heterogeneity magnitudes. However, it may not provide immediate guidance in specifying a prior for the heterogeneity parameter for practical application.
  On the other hand, Rhodes \emph{et~al.}\citep{RhodesEtAl2015} and Turner \emph{et~al.}\citep{TurnerEtAl2015} fitted comprehensive, fully Bayesian models with a focus on prediction, and with the aim of yielding immediately applicable prior specifications.  To this end, Turner \emph{et~al.}\citep{TurnerEtAl2015} utilized a log-normal model for binary outcomes,  
  while Rhodes \emph{et~al.},\citep{RhodesEtAl2015} after considering several alternatives, eventually settled on a log-Student\mbox{-}$t_5$ model for standardized mean differences.
  
  While the results from such previous analyses are immediately applicable for certain investigations, in some cases it is desirable to perform an analogous investigation based on a specific selected data base; an example scenario also sketched in the example application below is that of a health technology assessment (HTA) authority that may be interested in evaluating the heterogeneity commonly encountered in a specific set of past analyses.
  The technical problem essentially is that of \emph{a meta-analysis of heterogeneity estimates} with a focus on the heterogeneity ``population'' and prediction of a new (``future'') heterogeneity value. The heterogeneity estimates themselves each originate from a meta-analysis, and they are usually not well summarized simply by a point estimate and a standard error. The relevant data basis hence is a complete set of \emph{several} meta-analyses (i.e., their included effect estimates and standard errors); at least in the medical context, such original data are quite commonly reported as required by the \textsc{prisma}~state\-ment.\citep{LiberatiEtAl2009}
  
  In the present investigation, we formalize the approach originally established by Rhodes \emph{et~al.}\citep{RhodesEtAl2015} and Turner \emph{et~al.}\citep{TurnerEtAl2015} in a slightly more general way, including a range of parametric distribution families and suggestions on how to summarize the resulting posterior distributions for communication and application. 
  Modeling here includes choice of the random-effects distribution, and the focus is on prediction.
  While we showcase the approach using a freely available example data set, we also provide JAGS and \textsf{R}~code to facilitate application of the same approach to other data sets.
  
  The remainder of this article is structured as follows.  In section~\ref{sec:model}, we outline the problem in more detail and we specify the statistical model employed to investigate and infer heterogeneity distributions.  In section~\ref{sec:example}, the approach is showcased using a small example data set. Section~\ref{sec:discussion} then concludes with a discussion.

\section{Modeling heterogeneity}\label{sec:model}
  \subsection{Preliminary considerations}
    In the following, we consider the \emph{normal-normal hierarchical model (NNHM)}, in which estimation uncertainty as well as \emph{between-study heterogeneity} are accommodated using normal variance components. The magnitude of the heterogeneity component, which empirically manifests itself as excess variability beyond what could be attributed to uncertainty alone, scales with the standard deviation~$\tau$, which is generally unknown. In a Bayesian context, the specification of a prior distribution for~$\tau$ is then required.\citep{Roever2020,RoeverEtAl2021} In some instances, non-informative priors may be specified; in particular, in the context of a large number of studies included in a meta-analysis, varying the heterogeneity prior may be of little relevance to the resulting effect estimates. In general, however, consideration of prior information in the analysis will always be beneficial, and sometimes even necessary, e.g., when dealing with less that about 10~studies only, or when the focus is on computing marginal likelihoods. Prior distributions then may or may not involve empirical information.\citep{RoeverEtAl2021}

    In order to support an analysis by informing the prior choice using empirical data, first of all relevant, somehow ``representative'' data need to be identified.  \emph{How} exactly to arrive at such a data set is beyond the scope of the present investigation; here we assume an appropriate set of meta-analyses based on suitable effect measures to be given.
    In case one set out to gather a set of relevant heterogeneity estimates (and the associated meta-analyses), some guidance is given by the recommendations issued for meta-analyses in general, e.g., the \textsc{prisma} statement.\citep{LiberatiEtAl2009} Inclusion criteria should be defined, and restrictions might relate to characteristics like outcome (e.g., mortality), effect measure (e.g., odds ratio) or comparators (e.g., pharmacological vs. placebo), as was also done in the investigations by Rhodes \emph{et~al.}\citep{RhodesEtAl2015} and Turner \emph{et~al.}\citep{TurnerEtAl2015} An industrial investigator might consider a subset of their own sponsored trials, or an HTA institution might consider their own meta-analyses, and one might want to include time constraints (e.g., the past decade); either way, a transparent specification of the selection process will (as usual) make the investigation more convincing. In contrast to conventional systematic reviews (searches for \emph{studies} to be pooled), the search for meta-analyses would usually need to have a wider scope and may commonly include different indications, different treatments, different controls, etc.
    In the following, we discuss how such data may then be modelled and considered in a future analysis.

    It should be noted, however, that empirical data still remain only one aspect of several to guide prior specification; consideration of empirical data may in fact also be seen as ``complementary'' to other prior considerations.
    The heterogeneity parameter refers to the amount of variability \emph{between studies}, and, based on the context and the endpoint considered, it is usually possible to provide a rough specification of what amounts are considered likely, or from which point on the heterogeneity is deemed unreasonably large.\citep{RoeverEtAl2021}
    An ``empirical'' prior distribution might also serve as a ``sanity check'', e.g., in order to confirm whether some given prior specification appears too optimistic or otherwise unrealistic.  Furthermore, it will usually be possible to adapt a given prior distribution to make it ``less informative'', ``more conservative'' or ``more robust'' in a certain sense.\citep{RoeverEtAl2021}

    If no concrete (e.g., empirical) information is available, one can usually still constrain the plausible magnitude of the heterogeneity to be expected, leading to \emph{weakly informative} priors.  Such priors aim to facilitate analysis by bounding the heterogeneity in a conservative way; such specifications are also closely related to regularisation approaches.\citep{Gelman2009,ColeChuGreenland2013} Weakly informative priors for the heterogeneity parameter have been explored previously, and in particular their scaling depends on the context and the endpoint in question.\citep{RoeverEtAl2021,WilliamsEtAl2018}

  \subsection{Extending the NNHM}
    In order to infer a heterogeneity prior for use within the NNHM context, we specify an extension of the common NNHM allowing to accommodate data from \emph{several} meta-analyses at once, and including an additional model stage for the distribution of the associated heterogeneity parameters across meta-analyses. Suppose the external data consist of a collection of~$N$ meta-analyses, each of an individual size~$k_j$ ($j=1,\ldots,N$), and providing a set of $k_j$ estimates~$y_{ij}$ along with their associated standard errors~$\sigma_{ij}$ ($i=1,\ldots,k_j$). Depending on the effect measure considered, $y_{ij}$ and~$\sigma_{ij}$ (as well as the~$\mu_j$ and $\tau_j$ introduced below) are all expressed in the same units.

    The eventual aim of the analysis will be to determine the \emph{(posterior) predictive distribution} of a ``future'' heterogeneity value~($\tau^\star$); this distribution will be relevant as a heterogeneity prior for a subsequent random-effects meta-analysis.\citep{RhodesEtAl2015,TurnerEtAl2015,SchmidliEtAl2014}

  \subsection{The normal-normal stage}
    Each meta-analysis has an underlying true effect~$\mu_j$ and a heterogeneity~$\tau_j$ associated.  The $j$th meta-analysis is modelled via the ``usual'' NNHM:
    \begin{eqnarray}
      y_{ij} \,|\, \mu_j,\tau_j & \sim & \normaldistn(\mu_j, \, \sigma_{ij}^2 + \tau_j^2) \qquad (i=1,\ldots,k_j)\mbox{.}
    \end{eqnarray}
    Inference on different meta-analyses' effect parameters $\mu_j$ is effectively stratified by assuming a vague prior
    \begin{eqnarray}
      \mu_j \,|\, \mu_\mathrm{p},\sigma_\mathrm{p} & \sim & \normaldistn(\mu_\mathrm{p}, \, \sigma_\mathrm{p}^2)
    \end{eqnarray}
    for some ``neutral'' prior mean~$\mu_\mathrm{p}$ and some ``large'' (uninformative) prior standard deviation~$\sigma_\mathrm{p}$.\citep{Roever2020}

  \subsection{The heterogeneity stage}
    If one now chose a common ``noninformative'' prior for~$\tau$ across all $N$~analyses, one would essentially recover the posteriors that would also result from independent, separate analyses of all meta-analyses individually. Instead, we specify an additional model stage by defining a joint parametric (``prior'') distribution for the heterogeneity, whose (hyper-) parameters again are to be learned from the set of meta-analyses.
    The purpose of this additional model stage is to capture the distribution of the heterogeneity (standard deviation\mbox{-}) parameters across the set of $N$~meta-analyses.

    For the heterogeneity parameters, a common parametric distribution is assumed, whose parameters in turn are to be estimated. In general, we may express this as
    \begin{eqnarray}\label{eqn:taudistn01}
      \tau_j \,|\, \theta & \sim & P(\theta)
    \end{eqnarray}
    where the heterogeneity distribution's parameter(s)~$\theta$ are assigned another hyperprior
    \begin{eqnarray}\label{eqn:tauhyperprior01}
      \theta & \sim & H \mbox{.}
    \end{eqnarray}
    The \emph{predictive distribution} of a ``new'' heterogeneity value (e.g., in a future meta-analysis) is again defined through the conditional expression
    \begin{eqnarray}
      \tau^\star \,|\, \theta & \sim & P(\theta)\mbox{;}
    \end{eqnarray}
    its (prior or posterior) marginal distribution again results from marginalizing over the corresponding distribution of~$\theta$.

    To make the specification in~(\ref{eqn:taudistn01}) and~(\ref{eqn:tauhyperprior01}) more concrete, a common setting in practice may be to assume a half-normal distribution for~$\tau_j$, i.e.,
    \begin{eqnarray}\label{eqn:taudistn02}
      \tau_j \,|\, s & \sim & \halfnormaldistn(s)
    \end{eqnarray}
    where the common scale parameter~$s$ is unknown and is assigned another hyperprior, e.g.,
    \begin{eqnarray}\label{eqn:tauhyperprior02}
      s & \sim & \unifdistn(0,b)
    \end{eqnarray}
    for some fixed ``large'' upper bound~$b$ (e.g., a bound of $b=10$ should generally be appropriate for a log-OR as effect measure).

    In the following, inference first of all aims at inferring the common heterogeneity distribution ((\ref{eqn:taudistn01}) or (\ref{eqn:taudistn02}) above) and its parameters ($\theta$ or $s$ in the specifications above).  The relevant aspect for informing future meta-analyses then is the \emph{posterior predictive distribution} of a ``new'' heterogeneity value~$\tau^\star$.\citep{RhodesEtAl2015,TurnerEtAl2015} Here this means marginalizing over (\ref{eqn:taudistn02}) based on the scale parameter's posterior distribution. In the above example, this posterior predictive distribution will hence be a half-normal scale mixture.\citep{Lindsay}

    Many other sensible distributions may be conceivable instead of the half-normal specification in~(\ref{eqn:taudistn02}); the half-normal distribution, however, is probably the most common one. Inclusion of a scale parameter in the distribution's parametrization is often also sensible.\citep{RoeverEtAl2021} Alternatives to the half-normal would e.g.\ be exponential, half-Cauchy or log-normal distributions. All these may be parameterized in terms of a scale parameter, the log-normal distribution in addition possesses a shape parameter. The use of gamma or inverse-gamma distributions is usually not considered appropriate.\citep{Gelman2006}
    The different prior distributions differ for example in their shapes, their tail behaviour, or their complexity (number of parameters). For some guidance on their properties see e.g. the extensive discussion by R\"{o}ver \emph{et~al.} (2021).\citep{RoeverEtAl2021} Extensions of the model's heterogeneity stage may also be conceivable, e.g. in order to accommodate categorical covariables or time trends.

  \subsection{Inference}
    Inference on all unknowns in the model is easily implemented via Markov chain Monte Carlo (MCMC) methods,\citep{BDA3rd} e.g. utilizing JAGS\@.\citep{Plummer2003,R:rjags}  The parameters of primary interest here are the heterogeneity distribution's parameters ($\theta$) and the predictive distribution ($\tau^\star$). Alternative model specifications (different heterogeneity distributions in~(\ref{eqn:taudistn01}) or~(\ref{eqn:taudistn02})) may be investigated, and a formal model comparison may be implemented based on information criteria (such as the deviance information criterion (DIC);\citep{SpiegelhalterEtAl2002} see also the example below).

  \subsection{Summary and transfer}\label{sec:SumTransf}
    For any practical purpose, the ``plain'' MCMC samples alone (of~$\theta$ or~$\tau^\star$) would only be of very limited use. In order to communicate and apply the predictive distribution as a prior in a subsequent analysis, the MCMC samples would ideally be ``condensed'' into a simple parametric probability distribution.  There are three obvious options:
    \begin{enumerate}
      \item Derive a point estimate ($\hat{\theta}$), and then use a single, ``fixed'' instance of~(\ref{eqn:taudistn01}) (i.e., the conditional distribution~$p(\tau^\star|\hat{\theta})$) as an approximation to the predictive distribution for later application. The point estimate here may also be chosen ``conservatively'' (e.g., for a scale parameter, one may prefer the mean over the median, as this will usually be larger, or one might also use some upper posterior quantile in
        order to be on the safe side).
      \item Consider the parameter's (or parameters') uncertainty in addition. The predictive distribution results as a mixture distribution (e.g., a \emph{scale mixture} in case of a scale parameter~$\theta$). This may sometimes be approximated analytically based on the parameter's distribution's shape.\citep{RoeverEtAl2021}
      \item Fit a distribution (e.g., by matching moments or ML~fitting) to approximate the predictive distribution (of~$\tau^\star$) directly. Obvious choices for the distribution family may again be the heterogeneity model~(\ref{eqn:taudistn01}), or scale mixtures (or otherwise overdispersed versions) thereof.
    \end{enumerate}
    In the following, we will consider these three approaches for practical application. Note that the third approach is essentially the one also implemented by Rhodes \emph{et~al.}\citep{RhodesEtAl2015} and Turner \emph{et~al.},\citep{TurnerEtAl2015} or, in a related context, by Weber \emph{et~al.}\citep{WeberEtAl2019}

    When summarizing a posterior predictive distribution for practical application, the result will necessarily be a simplification, and a simple distributional form (a common, simple parametric distribution, with parameters rounded to reasonable accuracy) will foster applicability. On the other hand, care must be taken not to \emph{oversimplify} matters, and whenever deviations are accepted, one may aim to ensure that discrepancies are on the ``conservative'' or ``less informative'' side. Generally speaking, \emph{underestimation} of heterogeneity is commonly considered more harmful than \emph{overestimation}, and consequently one may try and preferably push potential bias towards larger heterogeneity.\citep{RoeverEtAl2021} In that sense, e.g., a stochastically larger distribution, a larger scale parameter, or a heavier-tailed distribution may be considered \emph{more conservative}.\citep{Roever2020,RoeverEtAl2021}

\section{Example application}\label{sec:example}
  \subsection{Example data set}
    As an example, we will utilize the meta-analysis data set compiled by Seide \emph{et~al.}\citep{SeideEtAl2018b,SeideEtAl2019} This data set contains data from 40~meta-analyses and includes a total of 131~studies. The data were originally assembled as a realistic test bed for comparing different meta-analysis methods for binary data. Reports published by the German Institute for Quality and Efficiency in Health Care (\emph{Institut f\"{u}r Qualit\"{a}t und Wirtschaftlichkeit im Gesundheitswesen, IQWiG)} were individually screened, and data from the first encountered  meta-analysis utilizing a binary endpoint and not including zero counts were extracted.\citep{SeideEtAl2018b,SeideEtAl2019}  In the following, we will consider logarithmic odds ratios (log-ORs) as effect measures. In this data set, the individual meta-analyses may be considered independent, since all of them originate from different reports and there is no overlap among the included studies. One might then consider the derived predictive distribution relevant 
    as a prior distribution
    for inference in a ``future'' report also employing a binary endpoint.

    Selection of this particular data set was somewhat opportunistic, as it was readily available, realistic and of a suitable size. However, one could easily imagine a very similar situation to come up in practice, where some health technology assessment (HTA) authority would be interested in characterizing the amount of heterogeneity encountered in a more or less confined subset of past analyses. In fact, an investigation of this kind is currently under way at the IQWiG; for more detail, see also Sec.~\ref{sec:discussion}.

    Before considering the empirical evidence in the following, it may be worthwhile reviewing what the corresponding \emph{a~priori} expectations may be. Here it is important to note that the endpoint (log-OR) is defined on a logarithmic scale, so that the additive variation implied by the between-study heterogeneity translates to a \emph{multiplicative} offset for the associated ORs. For example, if the heterogeneity was at $\tau=1.0$, this would imply study-specific effects within a range of mostly $\mu \pm 1.96$ (with 95\% probability) on the log-OR scale. On the OR~scale, this corresponds to effects within $\frac{\exp(\mu)}{7.1}$ and $\exp(\mu) \times 7.1$, potentially implying quite dramatic differences in effects between studies.\citep{SpiegelhalterEtAl} By considering these effect scales, one would commonly assume that $\tau$ would tend to be $<1.0$, e.g. by implementing a half-normal distribution with scale $0.18$,\citep{PrevostEtAl2000}, $0.32$,\citep{DiasEtAl2013c} or $0.50$\citep{FriedeRoeverWandelNeuenschwander2017a}.
    A half-normal prior distribution with scale~$0.50$ is commonly considered a conservative choice for endpoints such as log-ORs, as first of all it rules out unreasonably large heterogeneity ranges.\citep{FriedeRoeverWandelNeuenschwander2017a,RoeverEtAl2021,BenderEtAl2018}
    
  \subsection{The analysis}
\begin{figure*}[t]
      \centering
      {\includegraphics[width=0.80\linewidth]{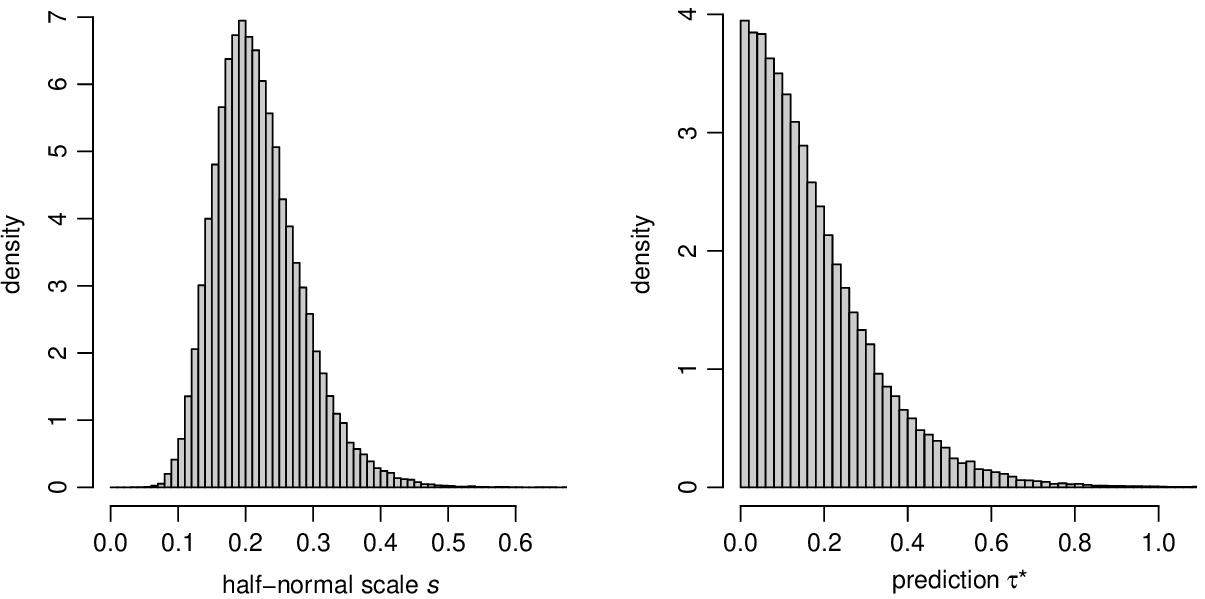}}
      \caption{\label{fig:histo01} Histograms of the posterior distribution of the half-normal scale parameter~$s$ (left panel) and of predictions~$\tau^\star$ (right panel).}
    \end{figure*}

    The model described in Section~\ref{sec:model}, using a half-normal distribution for the heterogeneity~$\tau$ (as in~(\ref{eqn:taudistn02})), a uniform prior for the scale parameter~$s$ (as in~(\ref{eqn:tauhyperprior02})) and a $\normaldistn(0,100^2)$~prior for the study-specific means was fitted using JAGS\@.\citep{Plummer2003,R:rjags}  
    The JAGS code implementing the model is given in Appendix~\ref{sec:JagsAppendix}.
    The resulting posterior distribution of the scale parameter~$s$ and the predictive distribution of~$\tau^\star$ are shown in Figure~\ref{fig:histo01}.
    Both distributions are also summarized using some descriptive statistics in Table~\ref{tab:McmcSummary}.
    The distribution of primary relevance for a ``future'' analysis is the predictive distribution of~$\tau^\star$. Subsequent inference hence needs to aim at roughly capturing its shape and relevant features.

  \subsection{Point estimation}
    An obvious and simple way of characterizing the predictive distribution is by simply deriving a ``representative'' point estimate of the heterogeneity distribution's parameter(s) (here: the scale parameter~$s$) and consider the corresponding conditional distribution. In the present case, we for example have a posterior mean scale of $\overline{s}=0.22$ (see Table~\ref{tab:McmcSummary}). The corresponding \emph{conditional} half-normal distribution ($p(\tau^\star|s\!=\!0.22)$) is shown along with the original histogram as a red line in Figure~\ref{fig:histo02}.
    Alternatively, one might also pick a more conservative (or ``worst case'') point estimate, for example, the upper 95\% quantile of the scale parameter's posterior instead.

  \subsection{Consideration of estimation uncertainty}
    Due to the uncertainty in the scale parameter~$s$, the histogram of predictions~$\tau^\star$ results as a \emph{(scale) mixture distribution}\citep{Lindsay} that generally includes more extreme (both small and large) values than a half-normal distribution could accommodate. An obvious example of a (half-) normal scale mixture is the (half-) Student\mbox{-}$t$ distribution, so that this may also be a natural choice for approximating the predictive distribution. The Student\mbox{-}$t$ distribution arises as a scale mixture of a normal distribution whose scale parameter (the normal standard deviation) follows a \emph{scaled inverse $\chi$-distribution}.\citep{RoeverEtAl2021} By matching a scaled inverse $\chi$-distribution to the scale parameter's distribution, we may then derive a corresponding half-Student\mbox{-}$t$ distribution as an approximation to the predictive distribution.

\begin{figure}[t]
      \centering
      {\includegraphics[width=0.40\linewidth]{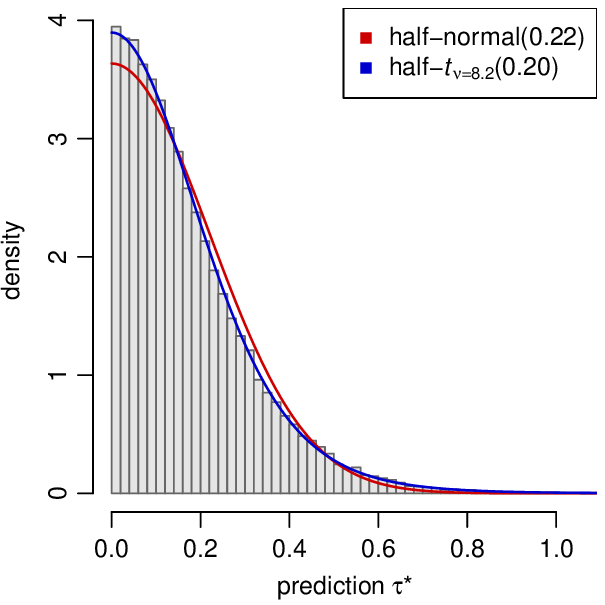}}
      \caption{\label{fig:histo02} Histogram of the heterogeneity predictions~$\tau^\star$ along with two approximations to its distribution.}
    \end{figure}

    In the present case, the scale parameter has a mean of~$0.22$ and standard deviation~$0.064$ (see Table~\ref{tab:McmcSummary}), corresponding to a coefficient of variation of~$0.29$. The resulting scale mixture may hence be approximated by a half-Student\mbox{-}$t$ distribution with $\nu=8.2$ degrees of freedom and scale $s=0.20$. This distribution is also shown in Figure~\ref{fig:histo02} as a blue line, and one can see that this yields an improved fit compared to the half-normal approximation. For details on fitting a Student\mbox{-}$t$ distribution as a normal scale mixture, see Appendix~\ref{sec:MixtureAppendix}.

    \begin{table}[t]
      \caption{\label{tab:McmcSummary}Summary statistics for the posterior distributions of scale parameter~$s$ and prediction~$\theta^\star$ (see also Figure~\ref{fig:histo01}).}
      \centering
      \begin{tabular}{lcc}
        \toprule
          & scale~$s$ & prediction~$\tau^\star$ \\
        \midrule
         mean               & 0.22\phantom{0}  & 0.17 \\
         standard deviation & 0.064            & 0.15 \\
         median             & 0.21\phantom{0}  & 0.14 \\
         95\% quantile      & 0.33\phantom{0}  & 0.46 \\
         99\% quantile      & 0.41\phantom{0}  & 0.66 \\
        \bottomrule
      \end{tabular}
    \end{table}

    \begin{table*}[b]
      \caption{\label{tab:McmcApprox}Summary statistics for the two
        approximations to the predictive distribution in comparison to
        the figures based on MCMC (see also Table~\ref{tab:McmcSummary}).}  \centering
      \begin{tabular}{lccccc}
        \toprule
         & & & \multicolumn{3}{c}{quantiles} \\
            \cmidrule(lr){4-6}         
         & mean & std.dev. & 50\% & 95\% & 99\% \\ 
        \midrule
         prediction~$\tau^\star$  & 0.17 & 0.15 & 0.14 & 0.46 & 0.66 \\[1ex]
         $\halfnormaldistn(0.22)$ & 0.18 & 0.13 & 0.15 & 0.43 & 0.57 \\
         half-Student\mbox{-}$t_{\nu=8.2}$(0.20)  & 0.18 & 0.15 & 0.14 & 0.46 & 0.67 \\
        \bottomrule
      \end{tabular}
    \end{table*}

  \subsection{Matching the predictive distribution}\label{sec:distributionMatching}
    Instead of matching the predictive distribution based on the scale parameter distribution's shape, the predictive distribution may also be fitted to the sampled~$\tau^\star$ values directly. Different distribution families may be reasonable here; more flexible (in particular: overdispersed) variations of the heterogeneity distribution (equations~(\ref{eqn:taudistn01}) or~(\ref{eqn:taudistn02})) are sensible choices.  In the present example, a half-Student\mbox{-}$t$ distribution hence remains an obvious choice for the parametric family.  Deriving maximum-likelihood estimates for degrees-of-freedom and scale parameters is straightforward using statistical software (via numerical optimization). Alternatively, one may also derive moment estimates; for the half-Student\mbox{-}$t$ distribution, one may first solve for the degrees-of-freedom and subsequently for the scale parameter (see Appendix~\ref{sec:StudenttAppendix} for more details). Another way to set up an overdispersed predictive distribution might be to fit a mixture distribution based on a small number of components, say 2--4.\citep{WeberEtAl2019}

    For the present example, both maximum-likelihood as well as moment estimation again yield estimates of~$8.2$ for the degrees-of-freedom~($\nu$), and of~$0.20$ for the scale~($s$) here; see also the corresponding distribution sketched in Figure~\ref{fig:histo02}. Table~\ref{tab:McmcApprox} also contrasts summary statistics of the predictive distribution (based on MCMC) to the half-normal or half-Student\mbox{-}$t$ fits.

  \subsection{Model selection}\label{sec:modelSelection}
    Beyond consideration of common conventions or simplicity, selection of a parametric model to be fitted to the samples may also be based on designated \emph{model selection} approaches.
    For example, the \emph{deviance information criterion (DIC)} may be computed from MCMC (JAGS) output in order to judge the goodness-of-fit or predictive performance of different distribution families.\citep{SpiegelhalterEtAl2002,GelmanHill,GelmanHwangVehtari2014,Meyer2016,Plummer2008} However, such model selection approaches are usually computationally challenging, and it may be questionable whether the often slight differences between a range of reasonable parametric distribution families are in fact of substantial practical relevance. For example, sensitivity analyses shown by R\"{o}ver \emph{et~al.}\citep{RoeverEtAl2021} seem to suggest that analyses based on priors from different families yet with matching medians may commonly lead to barely distinguishable results; this also seems to be confirmed by the alternative analyses shown in Section~\ref{sec:exampleVariations}.

    \begin{table}[t]
      \caption{\label{tab:DICs}DIC~values for the comparison of three alternative models based on the example data. A lower DIC~value indicates a better model fit. The corresponding predictive distributions are also summarized (in analogy to Table~\ref{tab:McmcApprox}).}  \centering
      \begin{tabular}{lccccccc}
        \toprule
          & & \multicolumn{5}{c}{predictive distribution ($\tau^\star$)} \\
          \cmidrule(lr){3-7}
          model       & DIC   & mean & st.dev. & 50\% & 95\% & 99\% \\
        \midrule
          half-normal & 163.8 & 0.17 & 0.15 & 0.14 & 0.46 & 0.66 \\
          exponential & 167.7 & 0.17 & 0.19 & 0.11 & 0.54 & 0.91 \\
          log-normal  & 178.0 & 0.26 & 3.46 & 0.11 & 0.61 & 1.90 \\
          half-Cauchy & 212.8 &      &      & 0.08 & 1.10 & 5.46 \\
        \bottomrule
      \end{tabular}
    \end{table}

    In the present example, we may compare a half-normal model with  analogous ones utilizing an exponential, a half-Cauchy or a log-normal distribution for the heterogeneity (instead of the specification in (\ref{eqn:taudistn02})).  Table~\ref{tab:DICs} shows DIC values based on the example data; the half-normal model fits best here.
    
    \begin{table*}[t]
      \caption{\label{tab:modelFits}Parametric fits to the distributions shown in  Table~\ref{tab:DICs}. The approximations for the exponential, log-normal and half-Cauchy models are based on maximum-likelihood estimates.}  \centering
      \begin{tabular}{llcccccc}
        \toprule
          & \multicolumn{6}{c}{predictive distribution ($\tau^\star$)} \\
          \cmidrule(lr){2-7}
          model       & approximation & mean & std.dev. & 50\% & 95\% & 99\% \\
        \midrule
          half-normal & $\halfnormaldistn(0.22)$       & 0.18 & 0.13 & 0.15 & 0.43 & 0.57 \\
          half-normal & half-Student-$t$($8.2$; $0.20$)   & 0.18 & 0.15 & 0.14 & 0.46 & 0.66 \\
          exponential & Lomax($9.9$; $1.5$)       & 0.17 & 0.19 & 0.11 & 0.53 & 0.89 \\
          log-normal  & log-Normal($-2.6$; $1.7$) & 0.32 & 1.30 & 0.07 & 1.22 & 3.88 \\
          half-Cauchy & half-Cauchy($0.10$)       &      &      & 0.10 & 1.27 & 6.37 \\
        \bottomrule
      \end{tabular}
    \end{table*}

  \subsection{Considerations beyond empirical data}
    Empirical data will only ever be one among several aspects in the specification of a prior distribution.\citep{Jaynes1968,KassWasserman1996,Gelman1996}  Other aspects include operating characteristics and robustness,\citep{SchmidliEtAl2014,OHaganPericchi2012} considerations of conservatism properties,\citep{Roever2020} features like lower-tail and upper-tail behaviours,\citep{RoeverEtAl2021} or simplicity\citep{JefferysBerger1992}
    (see also Sec.~\ref{sec:SumTransf}).
    In the present example, if simplicity is desired, the half-normal approximation may be preferred to the Student\mbox{-}$t$ model. If on the other hand there is some doubt about the direct relevance of the empirical data to the given meta-analysis problem, a ``more conservative'' distribution (e.g., a greater scale parameter) might be specified.\citep{Roever2020}.
    
    Another way to ``robustify'' a prior, or to reconcile several prior information sources, is by implementing a \emph{mixture prior}. The basic idea is that the prior information may be composed of several (mutually exclusive) components, for example, besides the hypothesis that the historical data are of immediate relevance to the present analysis, one may want to also consider the alternative that it is unrelated and hence some vague prior would apply. Both the ``informative'' and ``vague'' priors may be combined by attaching probabilities to both options and using a two-component mixture distribution of both as the eventual prior.\citep{SchmidliEtAl2014,RoeverWandelFriede2018} Inference would eventually also consider to what extent the present data appear consistent with one or another component; the effect is greater conservatism as well as a gain in \emph{robustness} to discrepancies between the historical and current data.

  \subsection{Example application}\label{sec:appli}
    We continued the original search through IQWiG publications to find the subsequent ``41st'' qualifying meta-analysis to apply the derived prior on.  
    \begin{figure*}[t]
      \centering
      {\includegraphics[width=0.95\linewidth]{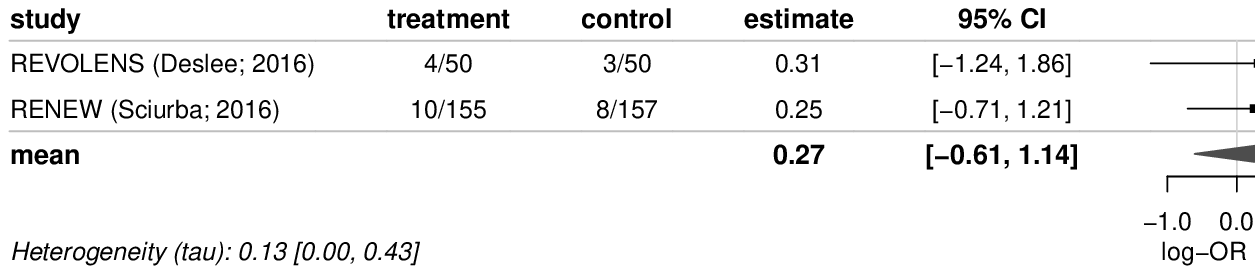}}
      \caption{\label{fig:study41a}Forest plot illustrating the meta-analysis of log-OR estimates from a subsequent IQWiG report\citep{IQWIG2017a} based on the heterogeneity prior derived from previous, related meta-analyses (a half-Student\mbox{-}$t$ distribution with $8.2$~degrees of freedom and scale~$0.20$).}
    \end{figure*}
    Report~\emph{\mbox{N14-04}}\citep{IQWIG2017a} (Fig.~32, p.~233) reports on a meta-analysis involving two studies, and comparing \emph{endobronchial coil} vs.\ \emph{no therapy} for lung volume reduction in severe pulmonary emphysema.\citep{DesleeEtAl2016,SciurbaEtAl2016,GarnerShah2020} The endpoint considered here is the log-OR of \emph{overall mortality at 12~months}. The relevant data are illustrated in Figure~\ref{fig:study41a}.  The two studies yield very similar log-OR estimates, while the standard errors differ. The (DL)~heterogeneity estimate turns out as zero in this case, so that the original (frequentist) analysis effectively was based on a ``common-effect'' estimate of a log-OR of $0.27$ [$-0.55$, $1.08$].\citep{IQWIG2017a}
    
    Given that the half-normal model provides the best fit to the historical data (see Table~\ref{tab:DICs}) and that the half-Student\mbox{-}$t$ yields a slightly better approximation to the predictive distribution than the half-normal (see Figure~\ref{fig:histo02}), here we select the half-Student\mbox{-}$t_{\nu=8.2}(0.20)$ approximation for analysis.
    Some properties of this distribution are sketched in Figure~\ref{fig:histo02} or in Table~\ref{tab:McmcApprox}.
    The resulting pooled effect estimate is shown in Figure~\ref{fig:study41a}. The resulting credible interval is about 8\% wider than the originally reported frequentist interval due to the increased heterogeneity incorporated in the Bayesian model. The heterogeneity posterior barely differs from its prior, as one can also see when comparing the heterogeneity estimate shown in the bottom left of Figure~\ref{fig:study41a} with the prior quantiles shown in Table~\ref{tab:McmcApprox} (prior and posterior densities are also shown side by side in Figure~\ref{fig:study41c} in the Appendix). 

\subsection{Variations of the example analysis}\label{sec:exampleVariations}
  \subsubsection{Alternative analyses}
    For comparison, we also analyzed the same data using several alternative approaches, including the alternative priors mentioned in Sec.~\ref{sec:modelSelection} that were derived based on differing distribution families for the heterogeneity, a Bayesian approach with a common weakly informative prior, as well as three frequentist estimators.
    \begin{figure*}[b]
      \centering
      {\includegraphics[width=0.99\linewidth]{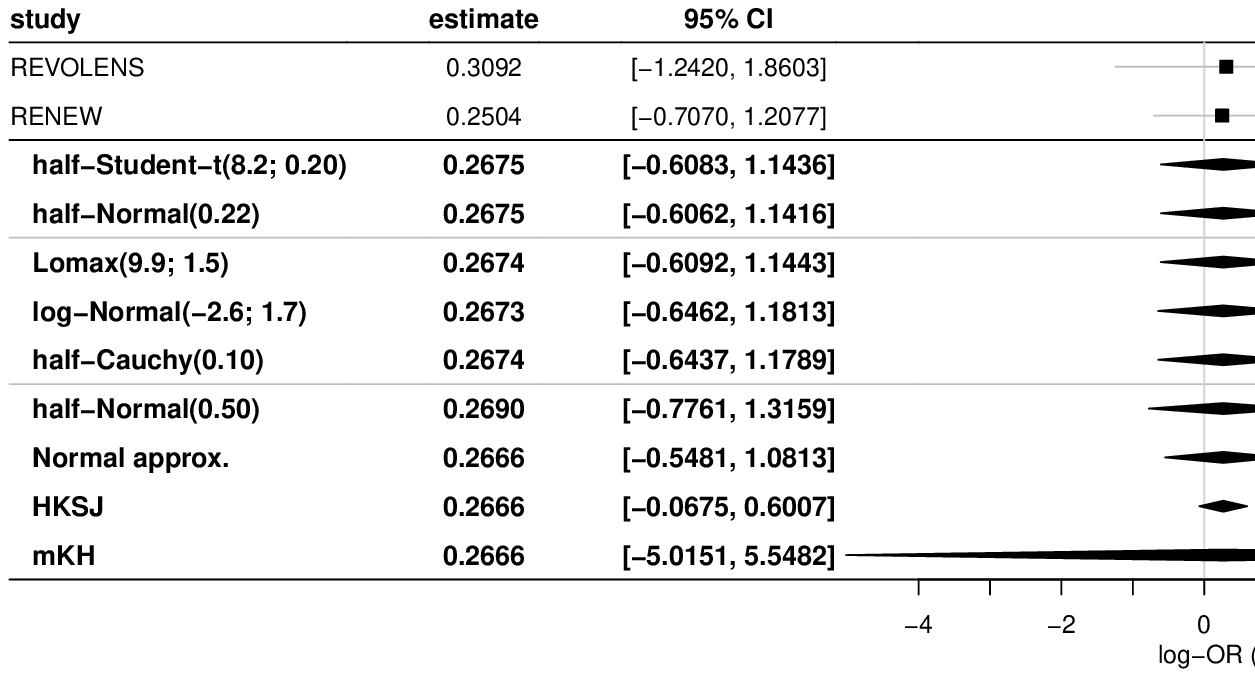}}
      \caption{\label{fig:study41b}Forest plot illustrating
        different (Bayesian and frequentist) analyses of the example data from
        Section~\ref{sec:appli}.}
    \end{figure*}
    
    Figure~\ref{fig:study41b} illustrates the different results jointly in a forest plot.
    The first two (half-Student\mbox{-}$t$ and half-normal) are the ones also shown in Figure~\ref{fig:histo02}, and are based on assuming a half-normal distribution for the random effects.
    The following three lines show the summary estimates resulting from assuming alternative models for the external data (see also Table~\ref{tab:modelFits}).
    The last four lines show common analyses not considering the external data; the half-Normal heterogeneity prior with scale~0.50 is commonly considered \emph{weakly informative} for a log-OR endpoint, as in the present context.\citep{FriedeRoeverWandelNeuenschwander2017a,RoeverEtAl2021}
    The following three confidence intervals correspond to common frequentist analysis methods and are based on a simple normal approximation (which was also utilized in the original analysis\citep{IQWIG2017a}) and the Hartung-Knapp-Sidik-Jonkman (HKSJ) and the modified Knapp-Hartung (mKH) intervals,\citep{RoeverKnappFriede2015} which are both based on Student\mbox{-}$t$ quantiles.
    
    Use of the different approximations to the predictive distribution in the half-normal model (half-normal or half-Student\mbox{-}$t$; see Figure~\ref{fig:histo02}) barely makes a noticeable difference for the eventual analysis.
    Assuming alternative random-effect distributions instead of the half-normal model leads to slightly differing predictive distributions as heterogeneity priors (see also Section~\ref{sec:modelSelection} and Table~\ref{tab:DICs}).
    Despite the differences apparent in the predictive distributions (Table~\ref{tab:DICs}) and their corresponding parametric approximations (Table~\ref{tab:modelFits}), all five resulting meta-analytic estimates still turn out very similar here, so one might pragmatically choose the simple half-normal prior here.
    The Bayesian estimate based on the weakly informative prior is more conservative, resulting in a wider interval (e.g., 20\% wider than for the half-Normal(0.22) prior).
    
    The different frequentist approaches suffer from the small sample size of only $k=2$ studies and appear to be either too optimistic or overly conservative. Since the heterogeneity point estimate is zero here, the frequentist intervals effectively correspond to ``common-effect'' analyses in this case. The HKSJ interval turns out roughly half as wide as the Normal interval, which may be considered counterintuitive.\citep{RoeverKnappFriede2015} The mKH interval on the other hand is extremely conservative (about six times as wide as the Normal interval), which again is not so uncommon for a meta-analysis of two studies only.\citep{FriedeRoeverWandelNeuenschwander2017b,BenderEtAl2018}

  \subsubsection{Differing amounts of historical data}
    When embarking on an investigation of ``historical'' meta-analyses as in the present example, an obvious question is \emph{how many} meta-analyses would be required or sufficient. This is probably hard to answer in any generality, as it also very much depends on the sizes~($k_j$) of involved meta-analyses, but we may at least shed some light by investigating variations of the example discussed above. To this end, besides the ``full'' set of $N=40$ studies, we restrict the data to smaller subsets including the most recent $20$, $10$ or $5$ studies.
    Table~\ref{tab:exampleSubsets} shows the resulting predictive distributions and corresponding approximations when considering increasingly smaller subsets of the total of 40~meta-analyses. One can see that when considering fewer heterogeneity estimates, the restricted data essentially are not able to ``rule out'' larger heterogeneity ranges, and so the resulting predictive distributions have larger means, medians and other quantiles. From the relative magnitude of mean and standard deviation (and the corresponding coefficient of variation) as well as from the Student\mbox{-}$t$ distributions' degrees-of-freedom parameters, one can see that the predictive distributions also become increasingly heavy-tailed. The cases where the degrees-of-freedom parameter is large are those where a half-normal approximation may fit comparably well; when the degrees-of-freedom parameter is low, the half-Student\mbox{-}$t$ distribution may be substantially more accurate.
    
    \begin{table*}[h]
      \caption{\label{tab:exampleSubsets}Predictive distributions (and their approximations) based on analyses of data subsets.}  \centering
      \begin{tabular}{lcccccll}
        \toprule
          & \multicolumn{5}{c}{predictive distribution ($\tau^\star$)} & & \\
          \cmidrule(lr){2-6}
          data & mean & std.dev. & 50\% & 95\% & 99\% & \multicolumn{2}{c}{approximations} \\
        \midrule
          all $40$ analyses         & 0.17 & 0.15 & 0.14 & 0.46 & 0.66 & half-normal(0.22) & half-Student-$t$(8.2; 0.20) \\
          $20$ most recent analyses & 0.25 & 0.24 & 0.18 & 0.71 & 1.13 & half-normal(0.31) & half-Student-$t$(4.3; 0.26) \\
          $10$ most recent analyses & 0.45 & 0.45 & 0.32 & 1.32 & 2.17 & half-normal(0.56) & half-Student-$t$(3.8; 0.44) \\
          $5$ most recent analyses  & 0.65 & 0.90 & 0.38 & 2.14 & 4.28 & half-normal(0.82) & half-Student-$t$(2.7; 0.56) \\
        \bottomrule
      \end{tabular}
    \end{table*}
    
    It is also worth noting that for few analyses (only~10 or~5 included), the corresponding half-normal approximation conveys very little information --- a half-normal prior with scale~0.5 is commonly already seen as a very conservative choice in this context,\citep{FriedeRoeverWandelNeuenschwander2017a,RoeverEtAl2021} and so the subsets of only~5 or~10 meta-analyses, corresponding to half-normal priors with larger scale, do not seem to add information beyond what may be assumed given already.
    Strictly speaking, such a-priori information could also be implemented in the hyperprior (see equations (\ref{eqn:tauhyperprior01}) or (\ref{eqn:tauhyperprior02}) in the analysis, which should then result in an implicit lower bound on the informativeness of the resulting predictive distribution. However, it may also be of interest to consider the different information sources in separation.

\section{Discussion}\label{sec:discussion}
  In this paper, we described a method allowing to translate historical meta-analysis data into a prior distribution for the heterogeneity parameter in a subsequent meta-analysis. The approach may serve to
  quantify plausible ranges for a heterogeneity parameter, or simply as a cross-check whether some given prior specification appears to be over-optimistic or too conservative relative to historical data.

  For illustration purposes, we considered the publicly available data set collected by Seide \emph{et~al.}\citep{SeideEtAl2018b,SeideEtAl2019} as a working example. The inferred predictive distribution may be somewhat realistic, but should not be understood as a recommendation of any generality. For example, while the data might be expected to be relatively homogeneous already, many of the analyses related to mortality endpoints, which where also found to be the least heterogeneous type of outcome in the related analysis of the Cochrane Database of Systematic Reviews (CDSR) by Turner \emph{et~al.}\citep{TurnerEtAl2015}
  We expect the type of endpoint considered to possibly play a role for the choice of prior. Endpoint-specific priors might be set up, whereas, in view of pragmatic considerations, it may also be possible to agree on a (simpler) common setting for different endpoints (e.g., odds ratios, relative risks, hazard ratios or standardized mean differences).

  The \textsf{R}~code to reproduce the analyses shown here is included in the online supplement. We intend to subsequently apply this method to a comprehensive collection of meta-analyses published by the German Institute for Quality and Efficiency in Health Care (\emph{Institut f\"{u}r Qualit\"{a}t und Wirtschaftlichkeit im Gesundheitswesen, IQWiG)} in order to systematically evaluate the empirical evidence on heterogeneity with respect to certain classes of effect measures or endpoints.
  The present results suggests that a half-normal distribution may serve the purpose well, so that the investigation would primarily yield estimates of half-normal scale parameters.
  These analyses could then form the basis for recommendations regarding the choice of prior in Bayesian random-effects meta-analyses for applications in health technology assessment (HTA).

  While initially an obvious approach might have been to consider collections of heterogeneity \emph{point estimates} for this purpose, it soon became obvious that these would only yield a very coarse indication of plausible heterogeneity ranges; a comparison of the predictive distribution and the distribution of point estimates for our example data is also illustrated in Appendix~\ref{sec:tauEstimateAppendix}.  We then set up a simple extension of the NNHM to accommodate a heterogeneity distribution and to infer the predictive distribution that would be useful for prior specification. 
  However, there may always be situations where heterogeneity estimates are available while the original studies are not;\citep{vanErpEtAl2017} in such cases, a collection of heterogeneity point estimates may still provide a reasonable approximation to the predictive distribution, as observed in the example. Overdispersion and bias of the heterogeneity estimates relative to the predictive distribution may be less of an issue when the original meta-analyses are reasonably large.
  Many model extensions or variations would be conceivable; for example, in view of binary endpoints, a binomial-normal hierarchical model similar to the one implemented by Turner \emph{et~al.}\citep{TurnerEtAl2015} (which would allow to dispose of the normal approximation at the first model stage), or a model accommodating subgroups of studies as utilized by Rhodes \emph{et~al.}\citep{RhodesEtAl2015} or Turner \emph{et~al.}\citep{TurnerEtAl2015}

  Although the heterogeneity prior plays an important role in the setup of a meta-analysis, many other details are at least equally relevant, such as the study selection, choice of estimand, effect measure or the statistical model. Empirical information will usually only constitute one of several aspects to inform or contribute to prior specification; further aspects to be considered include a prior's tail behaviour, or robustness or conservatism features.\citep{Roever2020,RoeverEtAl2021} While the availability of a coherent approach to accurately translate empirical data into a prior specification is convenient, we expect a rather rough summary of the resulting predictive distribution to commonly be sufficient.
  Unlike in usual systematic reviews, an exhaustive search may not be necessary and a smaller (but representative) sample may be sufficient.


\ack
\section*{Acknowledgment}
  Support from the \emph{Deutsche Forschungsgemeinschaft (DFG)} is
  gratefully acknowledged (grant number \mbox{FR~3070/3-1}).
  
\section*{Conflicts of interest}
  The authors have declared no conflict of interest.

\section*{Data availability}
  The data that supports the findings of this study are publicly available\citep{SeideEtAl2018b}, \textsf{R}~code allowing to reproduce calculations is provided in the supplementary material of this article.


\appendix
  \section{JAGS code}\label{sec:JagsAppendix}
    The code shown in Table~\ref{tab:JagsCode} defines the JAGS implementation of the model described in Section~\ref{sec:model}.\citep{Plummer2003,R:rjags} The data, a total of \texttt{N} effect estimates from \texttt{n} meta-analyses are provided in terms of the vectors of effect estimates~\texttt{y} and variances (squared standard errors)~\texttt{v}. Individual studies are allocated to meta-analyses via the vector of index variables~\texttt{id}.  The effects' vague prior is given through its mean \texttt{effectPriorMean} and standard deviation \texttt{effectPriorSD}.

    The model for the heterogeneity standard deviations~\texttt{tau} is eventually specified in the final loop; here this is a half-normal distribution with a scale parameter (\texttt{tauScale}).  The scale parameter is estimated from the data after specification of a vague prior, a uniform distribution with upper bound~\texttt{tauScalePriorMax}. Eventually, samples from the predictive distribution (\texttt{tauPrediction}) are also generated along the way.
    \begin{table*}[h]
      \caption{\label{tab:JagsCode}JAGS code implementing the model described in Appendix~\ref{sec:JagsAppendix}.}  
  \begin{verbatim}
model {
  # the normal-normal hierarchical model (NNHM):
  for (i in 1:N) { # loop over (N) individual studies:
    y[i] ~ dnorm(mu[id[i]], pow(v[i]+pow(tau[id[i]], 2), -1))
  }
  for (i in 1:n) { # loop over (n) meta-analyses:
    mu[i] ~ dnorm(effectPriorMean, pow(effectPriorSD, -2))
  }
  # the heterogeneity prior:
  for (i in 1:n) { # loop over (n) meta-analyses:
    # half-normal distribution for tau:
    tau[i] ~ dnorm(0.0, pow(tauScale, -2)) T(0,)
  }
  tauScale ~ dunif(0.0, tauScalePriorMax)
  tauPrediction ~ dnorm(0.0, pow(tauScale, -2)) T(0,)
}
\end{verbatim}
  \end{table*}

  \section{Scale mixture parametrization}\label{sec:MixtureAppendix}
    \subsection{Normal scale mixture}
      Suppose a half-normal distribution's scale parameter~$s$ has mean~$\mean(s)$, variance~$\vari(s)$ and coefficient of variation~$\cv(s)=\frac{\sqrt{\vari(s)}}{\mean(s)}$.  Then the resulting mixture distribution may be approximated by a half-Student\mbox{-}$t$ distribution by assuming that the scale followed a \emph{scaled inverse~$\chi$ distribution} with matching moments.\citep{RoeverEtAl2021}

      The matching half-Student\mbox{-}$t$ distribution's parameters (degrees-of-free\-dom $\nu_t$ and scale~$s_t$) then depend on $\cv(s)$ and $\mean(s)$. First, the degrees-of-freedom may be solved for numerically based on the coefficient of variation~$\cv(s)$ by equating the scaled inverse~$\chi$ distribution's coefficient of variation (which does not depend on its scale parameter) to match~$\cv(s)$. Then the half-Student\mbox{-}$t$ distribution's scale results as $s_t = \frac{\mean(s)}{E_{\scriptsize\mbox{Inv-$\chi$}}(\nu_t, \sqrt{\nu_t})}$, where $E_{\scriptsize\mbox{Inv-$\chi$}}(\nu_t, \sqrt{\nu_t})$ is the expectation of a scaled inverse~$\chi$ distribution with degrees-of-freedom~$\nu_t$ and scale~$\sqrt{\nu_t}$.\citep{RoeverEtAl2021}

    \subsection{Exponential scale mixture}
      Suppose an exponential distribution's scale (or \emph{inverse rate}) parameter~$s$ has mean~$\mean(s)$, variance~$\vari(s)$ and coefficient of variation~$\cv(s)=\frac{\sqrt{\vari(s)}}{\mean(s)}$.  Then the resulting mixture distribution may be approximated by a Lomax distribution by assuming that the scale followed an inverse gamma distribution with matching moments.\citep{RoeverEtAl2021}

      The Lomax distribution is parameterized in terms of shape~$\alpha$ and scale~$\lambda$, with expectation~$\frac{\lambda}{\alpha-1}$ and variance~$\frac{\lambda^2\alpha}{(\alpha-1)^2 (\alpha-2)}$. The corresponding parameter values follow from $\cv(s)$ and $\mean(s)$ and are given by $\alpha=2+\frac{1}{\cv(s)^2}$ and $\lambda=\mean(s)\bigl(1+\frac{1}{\cv(s)^2}\bigr)$.\citep{RoeverEtAl2021}

   \subsection{Log-normal distribution}
      Another obvious and common distribution for the heterogeneity~$\tau$ is the log-normal distribution, resulting from modeling the logarithmic heterogeneity parameter~($\log(\tau)$) based on location~($\mu$) and scale~($\sigma$). In the context of the other models discussed here, it is useful to consider an alternative, less common parametrization of the log-normal distribution, namely, based on the exponentiated mean parameter $\vartheta=\exp(\mu)$. Re-writing the log-normal density~$f(x)$ in this form, it becomes apparent that $\vartheta$~constitutes a \emph{scale parameter} here, while $\sigma$~is a \emph{shape parameter}:
      \begin{equation}
        f(x) \;=\; \frac{1}{\vartheta}\frac{1}{(x/\vartheta)\,\sigma\sqrt{2\pi}}
                   \,\exp\biggl(-\frac{\bigl(\log(x/\vartheta)\bigr)^2}{2\sigma^2}\biggr)\mbox{.}
      \end{equation}
      This allows to treat the log-normal distribution analogously to the other distributions discussed e.g.\ in Section~\ref{sec:model}.
      In this parametrization, the distribution is characterized by
      \begin{center}
      \begin{tabular}{ll}
        \toprule
          median             & $\vartheta$ \\
          mean               & $\vartheta \, \sqrt{\exp(\sigma^2)}$ \\
          variance           & $\vartheta^2 \, \exp(\sigma^2) \, \bigl(\exp(\sigma^2)-1\bigr)$ \\
          standard deviation & $\vartheta\, \sqrt{\exp(\sigma^2) \, \bigl(\exp(\sigma^2)-1\bigr)}$ \\
          coefficient of variation & $\sqrt{\exp(\sigma^2)-1}$\\
        \bottomrule
      \end{tabular}
      \end{center}

      When heterogeneity is modelled using a log-normal distribution in the extended NNHM (see Section~\ref{sec:model}), the same distribution may also be a reasonable choice for approximating the (usually more dispersed) predictive distribution (of~$\tau^\star$). While a \emph{scale mixture} of log-normal distributions does not have a simple analytic form, this may still be motivated by considering the logarithmic scale: implementing uncertainty in $\mu=\log(\vartheta)$ using an additive normal term again leads to a log-normal distribution (with a larger $\sigma$~parameter).  Strictly speaking, the predictive distribution for a normal distribution with unknown mean and variance is a Student\mbox{-}$t$ distribution,\citep{BDA3rd} however, fitting a three-parameter distribution may well be an exaggeration in the present context.

      Log-normal distributions have been used for modelling predictive distributions e.g. by Turner \emph{et~al.}\citep{TurnerEtAl2015} Similarly, Rhodes \emph{et~al.}\citep{RhodesEtAl2015} used a log-Student\mbox{-}$t$-distribution, albeit with a pre-specified, fixed degrees-of-freedom parameter ($\nu=5$).

  \section{Moment estimation for the half-Student-t distribution}\label{sec:StudenttAppendix}
    The first two moments (expectation and variance) of a half-Student\mbox{-}$t$ distribution with degrees-of-freedom~$\nu$ and scale~$\sigma$ are given by
    \begin{equation}
      \expect[X] \;=\; 2\sigma \sqrt{\frac{\nu}{\pi}} \, \frac{\Gamma(\frac{\nu+1}{2})}{\Gamma(\frac{\nu}{2}) \, (\nu-1)}
      \qquad \mbox{(for $\nu>1$)}
    \end{equation}
    and
    \begin{equation}
      \var(X) \;=\; \sigma^2 \Biggl(\frac{\nu}{\nu-2}-\frac{4\nu}{\pi\,(\nu-1)^2}\biggl(\frac{\Gamma(\frac{\nu+1}{2})}{\Gamma(\frac{\nu}{2})}\biggr)^2\Biggr)
      \qquad \mbox{(for $\nu>2$).\citep{PsarakisPanaretos1990}}
    \end{equation}
    Its coefficient of variation hence results as
    \begin{eqnarray}\label{eqn:studenttCV}
      \frac{\sqrt{\var(X)}}{\expect[X]} 
      &=&
      \sqrt{\frac{\pi\,(\nu-1)^2}{4\,(\nu-2)}\biggl(\frac{\Gamma(\frac{\nu}{2})}{\Gamma(\frac{\nu+1}{2})}\biggr)^2-1}
      \qquad \mbox{(for $\nu>2$),}
    \end{eqnarray}
    which is independent of the scale~$\sigma$ and $ > \sqrt{\frac{\pi}{2}-1}\approx 0.75$. A half-Student\mbox{-}$t$ distribution's moment estimates may hence be computed by first matching the coefficient of variation (given that it is greater than~$0.75$) to yield a degrees-of-freedom estimate, and subsequently matching the expectation for a scale estimate. Solving equation~(\ref{eqn:studenttCV}) for the degrees-of-freedom~$\nu$ may be done numerically.

    For example, suppose a coefficient of variation of~$0.8$ and an expectation of~$0.5$ were aimed for. Numerically solving equation~(\ref{eqn:studenttCV}) (e.g., utilizing \textsf{R}'s ``\texttt{uniroot()}'' function) yields a degrees-of-freedom estimate of $\hat{\nu}=13.6$. A half-Student\mbox{-}$t$ distribution with $\nu=13.6$ degrees of freedom (and scale~$1$) would have an expectation of~$0.85$. The scale's moment estimator then equals $\hat{\sigma}=\frac{0.5}{0.85}=0.59$, i.e., the half-Student\mbox{-}$t$ distribution with $\nu=13.6$ and $\sigma=0.59$ has an expectation of~$0.5$ and coefficient of variation~$0.8$.

  \section{Comparison to frequentist point estimates}\label{sec:tauEstimateAppendix}
    Since a heterogeneity \emph{estimate} is associated with additional variance due to estimation uncertainty, one would expect the distribution of estimates to be overdispersed relative to the distribution of true values.  
    \begin{figure}[h]
      \centering
      {\includegraphics[width=0.55\linewidth]{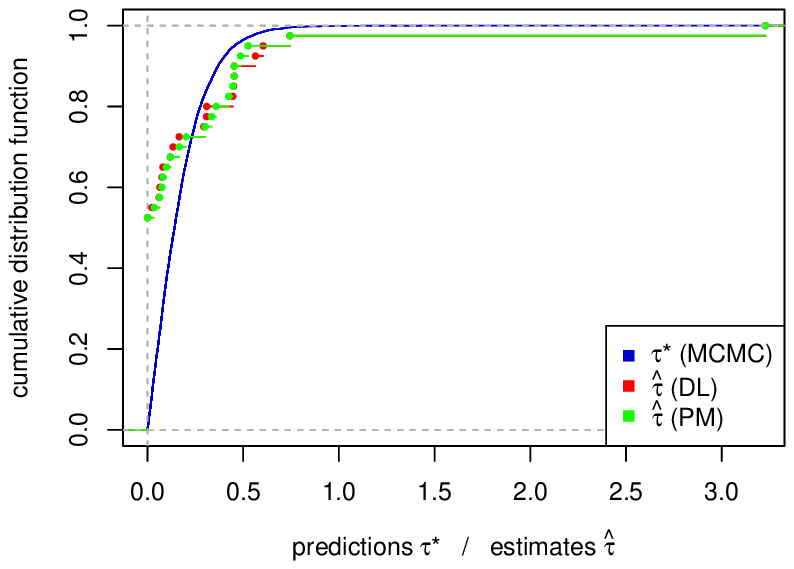}}
      \caption{\label{fig:PredEsti} Comparison of predictive distribution vs.\ the empirical distribution of 40~heterogeneity \emph{point estimates} for the example data set.}
    \end{figure}
    Figure~\ref{fig:PredEsti} contrasts the predictive distribution (of~$\tau^\star$, as also shown in Figure~\ref{fig:histo02}) with the distributions of 40~heterogeneity estimates from the example data. The DerSimonian-Laird (DL) and Paule-Mandel (PM) estimates behave similarly here, a seizable fraction turns out as zero, and the overall picture appears to confirm the expected behaviour; in particular, in becomes evident that consideration of a mere collection heterogeneity \emph{estimates} is of only limited use for prior specification. For example, the median $\hat{\tau}$~value would be at zero, while the average would be at 0.21 (for both DL and PM) and would to a large extent be dominated by a single outlier.

  \section{Analyses of example data}\label{sec:study41Appendix}
\begin{figure*}[h]
      \centering
      {\includegraphics[width=0.6\linewidth]{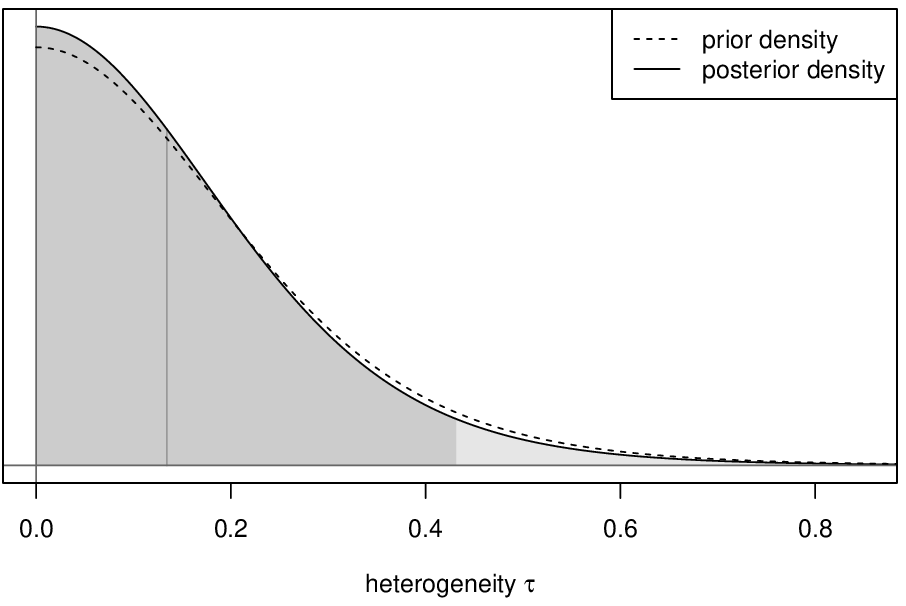}}
      \caption{\label{fig:study41c}Prior and posterior densities for the analysis shown in Figure~\ref{fig:study41a}. The vertical line indicates the posterior median, and the dark grey area shows the 95\% credible interval.}
    \end{figure*}

    Figure~\ref{fig:study41c} illustrates the prior and posterior densities corresponding to the analysis also shown in Figure~\ref{fig:study41a}. As to be expected, the posterior is very similar to the prior (a half-Student\mbox{-}$t$ distribution with $8.2$~degrees of freedom and scale~$0.20$) in this case of only $k=2$~studies.

\bibliography{literature}

\end{document}